\font\caps=cmcsc10 at 12pt
\font\eightrm=cmr8 at 8pt
\documentclass{oxarticle}
\usepackage{amsmath, amssymb}
\usepackage{feynmf}

\newcommand{\GaH}{Gauginos and Higgsinos}
\newcommand{\pgh}{{Section}}

\newcommand{\UC}{Un-Chiral}
\newcommand{\UCM}{Un-Chiral Multiplet}

\newcommand{\xp}{{\x\cdot\pa}}

\newcommand{\SG}{{Supergravity}}

\newcommand{\ccp}{{cosmological constant problem}}
\newcommand{\CTR}{{Exchange\;Transformation}}
\newcommand{\CHC}{{Complex\; Half-Chiral}}
\newcommand{\HC}{{Half-Chiral}}
\newcommand{\EP}{{Effective Potential}}
\newcommand{\CM}{{Chiral Multiplet}}

\newcommand{\HCM}{{Half-Chiral Multiplet}}

\newcommand{\eqm}{{equation of motion}}

\newcommand{\CT}{{Completion Term}}

\newcommand{\cY}{{\cal Y}}

\newcommand{\cA}{{\cal A}}

\newcommand{\cdss}{Chiral Dotted Spinor Superfield}

\newcommand{\CC}{Complex Conjugate}

\newcommand{\PB}{Master Equation}

\newcommand{\GSB}{Gauge Symmetry Breaking from \sg \ to \bsg }
\newcommand{\GSBS}{gauge symmetry breaking from \gto\ to $U(1)$}

\newcommand{\SM}{Standard Model}

\newcommand{\sg}{$SU(3) \times SU(2) \times U(1)$}

\newcommand{\bsg}{$SU(3)  \times U(1)$}

\newcommand{\gto}{$SU(2)  \times U(1)$}

\newcommand{\bt}{\begin{tabular}{c}}
\newcommand{\et}{\end{tabular}}

\newcommand{\eb}{\ee\be } 
\newcommand{\ebp}{\rt.\ee\be\lt.} 
\newcommand{\bmat}{\lt ( \begin{array} }
\newcommand{\emat}{  \end{array} \rt )}

\newcommand{\oH}{{\ov H}}

\newcommand{\oZ}{{\ov Z}}

\newcommand{\cP}{{\cal P}}

\newcommand{\oS}{{\ov S}}

\newcommand{\oG}{{\ov \G}}

\newcommand{\ED}{\end{document}}

\newcommand{\oC}{{\ov C}}

\newcommand{\cD}{{\cal D}}

\renewcommand{\a}{\alpha}	
\renewcommand{\b}{\beta}
\newcommand{\g}{\gamma}
\renewcommand{\d}{\delta}

\newcommand{\ve}{\varepsilon}

\newcommand{\h}{\eta}
\newcommand{\q}{\theta}

\newcommand{\x}{\xi}

\newcommand{\s}{\sigma}

\newcommand{\f}{\phi}
\renewcommand{\c}{\chi}

\newcommand{\w}{\omega}
\newcommand{\G}{\Gamma}
\newcommand{\D}{\Delta}

\newcommand{\X}{\Xi}

\newcommand{\W}{\Omega}
\newcommand{\la}{\label}
\newcommand{\ci}{\cite}

\newcommand{\ds}{\documentstyle}	
\newcommand{\fr}{\frac}

\newcommand{\pa}{\partial}
\newcommand{\ov}{\overline}
\newcommand{\be}{\begin{equation}}
\newcommand{\ee}{\end{equation}}
\newcommand{\ba}{\begin{array}} 
\newcommand{\ea}{\end{array}}
\newcommand{\bea}{\begin{eqnarray}}
\newcommand{\eea}{\end{eqnarray}}
\newcommand{\ra}{\rightarrow}
\newcommand{\Ra}{\Rightarrow}

\newcommand{\Lra}{\Leftrightarrow}
\newcommand{\lt}{\left}
\newcommand{\rt}{\right}

\newcommand{\ben}{\begin{enumerate}}
\newcommand{\een}{\end{enumerate}}
\newcommand{\bitem}{\begin{itemize}}
\newcommand{\eitem}{\end{itemize}}

\newcounter{orange} 
\newcommand{\articlenumber}{4934SSSMFINALX.tex}

\proofmodefalse
\usepackage{color} 
\begin{document}

\begin{center}

\vspace*{1in}

{ \huge The SSM with Suppressed SUSY Charge\\
[.5cm] }  

\vspace*{.1in}
%


\renewcommand{\thefootnote}{\fnsymbol{footnote}}

{\caps John A. Dixon\footnote{jadixg@gmail.com
} }\\[1cm] 

\end{center}

\normalsize
\vspace*{.2in}
 \begin{center}
 {\bf Abstract}
\end{center}

An earlier paper showed that it is possible to write down new  SUSY Actions in which it is not possible to define a Supersymmetry Charge.  SUSY is defined in these new Actions by the fact that they satisfy \PB s.
The new SUSY Actions are very easy to write down.  One simply takes a Chiral SUSY Action, coupled to Gauge and other Chiral Multiplets, and even \SG, if desired. Then one creates a new Action from this by exchanging all or part of the Scalar Field $S$ for a new Zinn Source $J$, and the corresponding part of the Zinn Source $\G$ for a new Antighost Field $\h$.  Since the original Action satisfies a \PB,  this exchange guarantees that the new Action will satisfy the new \PB. 
As was shown in the earlier paper, the new multiplets have fewer bosonic degrees of freedom than fermionic degrees of freedom. This is possible because they do not have a Supercharge. 

The resulting new SSM has no need for Squarks or Sleptons.  It does not need spontaneous breaking of SUSY, so that the cosmological constant problem does not arise (at least at tree level).  It mimics the usual non-supersymmetric Standard Model very well,  and the absence of large flavour changing neutral currents
is natural.  There is no need for a hidden sector, or a messenger sector, or explicit `soft' breaking of SUSY.  Spontaneous  \GSB\ implies the existence of two new very heavy Higgs Bosons with mass 13.4 TeV, slightly smaller than the energy of the LHC at 14 TeV. 
There is also a  curious set of  \GaH\  which have exactly the same masses as the Higgs and Gauge Bosons.  These do not couple to the Quarks and Leptons, except through the Higgs and Gauge Bosons.

\refstepcounter{orange}
{\bf \theorange}.\;
{\bf Introduction}  
This is the second of two papers on the topic of Chiral SUSY Theories with a Suppressed SUSY Charge.  The first paper was \ci{four}, and the results of  \ci{four}  are often referred to and used here.

 The Minimal Supersymmetric Standard Model (`MSSM') has been defined as the minimal Model based on SUSY such that it contains the smallest number of new particle states and new interactions consistent with phenomenology.  The `working hypothesis' here \ci{xerxes,haber,hepth} is usually that  all Multiplets are either Chiral or Gauge Multiplets \ci{west,
KBbook,
superspace,
WB,
ferrarabook,Weinberg3}, and that there is  spontaneous breaking of SUSY,
 with some unspecified, and presently undetected, SUSY Action.  That breaking is then manifested in the `visible' MSSM by explicit `soft' (usually massive) SUSY breaking terms added to the Action \ci{xerxes,haber,Weinberg3}.  The form of this breaking is  assumed to be communicated from the hidden sector to the visible sector by a messenger sector, and several different messenger ideas are currently used.  This Model has approximately 124 parameters, far more than the number that  were needed in the \SM. The main reason  for the invention of the hidden sector, and the messenger sector, and the large number of parameters, is that spontaneous breaking of SUSY implies sum rules for the masses after spontaneous SUSY breaking in the theory, and these are very hard to reconcile with experimental results, if they occur in the visible sector \ci{Weinberg3}.   
Another serious problem for this kind of SSM is that the spontaneous breaking of SUSY necessarily entails a fine tuning of the cosmological constant to one part in $10^{120}$, or other related problems,
when \SG\ is coupled \ci{weinbergcosmo,freedman}, or a huge vacuum energy  for the rigid theory before \SG\ is coupled.

The present paper will use a very different `working hypothesis' to construct a very different kind of SSM.  The working hypothesis here  is that we can 
replace any of the  \CM s in a starting Action with the new kinds of Multiplets that arise from the \CM s using the methods outlined in \ci{four}.  For each \CM, there is a choice of three different new Multiplets.  Two of these are  \HCM s and there is a third \UCM. 
In effect what happens is that we introduce certain \CTR s that remove a selected set of the original Scalar Fields  from the Action and replace them with Zinn Sources, while also replacing some of the original  Zinn Sources with Antighost Fields.  The  details of this are set out in \ci{four}.  As explained in \ci{four}, the  new \PB\ still yields zero for the One Particle Irreducible Vertex Functionals of the theory after this change, so supersymmetry is still present, though its consequences are changed in a major way, and in fact the 
Supercharge is gone from the modified parts. 

In this paper we use the \CTR s to remove all the Squarks and Sleptons from
the SSM, and also to remove half of the Scalar Fields in the three Higgs Multiplets that we start with here.  This means that the necessary parameters for the Model here are those of the \SM\footnote{These parameters are the 4 parameters of the CKM matrices, and six masses, for the six Quarks; and their analogue for the Leptons.  These are measured and tabulated for the Quarks in the Standard Model on page 214  of \ci{PDGstandardModel}.  For the Leptons see page 235 of \ci{PDGstandardModel}.} for the Matter Sector, plus five more real parameters for the Gauge/Higgs sector. Because the Sleptons and Squarks are not present, there is no  need for  spontaneous breaking of SUSY in the Matter sector.
No explicit breaking of SUSY, or hidden sector, or messenger sector are needed. The natural suppression of   flavour changing  neutral currents found in the \SM\ is restored in this new SSM.  In addition, the vacuum energy remains zero after \GSBS, which holds considerable promise for the \ccp\ set out in \ci{weinbergcosmo} and further explained in \ci{freedman}.

However, there are still mass degeneracies in the Higgs/Gauge sector, and more analysis is needed to determine whether the theory is viable with those in place.  The two new Higgs Bosons, with mass 13.4 TeV, predicted by the model, are probably beyond the practical range of current experimental tests, even though the maximum planned energy of the LHC is slightly larger at 14 TeV.

\refstepcounter{orange}
{\bf \theorange}.\;
{\bf Organization of this paper:}  
In this paper we will build what may be the simplest and minimal  SSM\ that can be made using the \CTR\ idea.  There are three \CM s for the Higgs sector.  We start with the usual pair\footnote{As is well known, one needs to have two doublets $(\fr{1}{2},\fr{1}{2})$ and $(\fr{1}{2},\fr{-1}{2})$ to give mass to the up and down quarks.}  of $SU(2) \times U(1)$ doublets with quantum numbers  $(\fr{1}{2},\fr{1}{2})$ and $(\fr{1}{2},\fr{-1}{2})$. Then we add one more Chiral Multiplet in the  $(1,0)$ representation  of  $SU(2) \times U(1)$.  With these three \CM s we can form two supersymmetric  gauge invariant mass terms and one supersymmetric gauge invariant Yukawa type term, assuming that the Action is of renormalizable type.  Including the $SU(2) \times U(1)$  coupling constants, we have five real parameters in the Gauge/Higgs sector.   

We start with the Gauge Higgs sector.  Then we perform \CTR s to remove half of the Scalars from each of these three \CM s, and we merge the two $(\fr{1}{2},\fr{1}{2})$ and $(\fr{1}{2},\fr{-1}{2})$ Multiplets by viewing them as  a Dirac Multiplet as described in \ci{four}.  These are   \HCM s.
Next all  Zinn Sources and the new Antighosts are set to zero.  This leaves a "Pure Field Action".
This Action has an \EP\ made from the remaining Scalars in the theory.  There is a nice solution for the Vacuum Expectation Values of those Scalars to implement \GSBS. 
 The masses of the Z Boson, the W Boson and the Higgs Boson, and the dimensionless electromagnetic and weak couplings, can be taken from the experimental values, and then the Model is totally determined (except for the Quark and Lepton sector, which is almost identical to the Standard Model, with parameters determined by experiment). 
One might think that this set of Higgs representations looks like it would generate more extra particles than are in the usual SSM.  But actually there are fewer particles, because the  \HCM s have only half the Bosonic degrees of freedom of a \CM.   
 As a result, the Model predicts that there is one more neutral Higgs and one charged Higgs, and it also predicts their masses.  Both of these are  very heavy, approximately 13,400 GeV= 13.4 TeV.  This analysis takes up most of the paper. 
 It shows that there are still supermultiplets, of a sort, but they have half of their  Scalars removed from the Chiral parts.

The reason that we choose to keep half the Scalars for the Higgs Multiplets is that the Scalars implement \GSBS\ without degeneracy between Higgs Scalar Bosons and Vector Bosons, which otherwise will happen, if one just uses \CM s for the Higgs.

The Gauge/Higgs part of the Action is introduced in \pgh s \ref{kinandzinnfogauge}
to \ref{VEVS}. The pure field (i.e. no Zinn Sources) Action can be obtained using the field shifts in 
\pgh\ \ref{shiftsdefe}.
Then all the mass terms are analyzed for the neutral 
 Gauge/Higgs sector in \pgh s 
\ref{neutghostsec}
to \ref{neutfermoion},
 and for the charged Gauge/Higgs sector in \pgh s 
\ref{startcharged}
to
\ref{endcharged}.
The parameters for this theory are obtained from the experimental results in \pgh\
\ref{numbers}.

A summary of the  Quark and Lepton action is in \pgh\
\ref{lepsandquarks}, and  the Gauge/Higgs sector is discussed a little more in \pgh\
\ref{gaugehiggsumm}.  
The remaining mass degeneracies and predictions and issues are discussed in
the Conclusion in \pgh\
\ref{conclusion}.

\refstepcounter{orange}\la{kinandzinnfogauge}
{\bf \theorange}.\;{\bf  Kinetic and Zinn
Actions for  Gauge Theories:}
We introduce the following Actions for the usual SUSY gauge theories.  The notation here was introduced in \ci{one}:

\be
\cA_{\rm Kinetic \;SU(2)}
=
 \int d^4 x \; \lt \{
-\c^{a\dot \a}
\cD^{ab}_{\a\dot\a}
\ov\c^{b \a}
-
\fr{1}{2}
F^a_{\dot \a\dot \b}
F^{a\dot \a\dot \b}
+
D^aD^a \rt \}
\la{kinchi}
\ee

\be
\cA_{\rm Kinetic \;U(1)}
=
 \int d^4 x \; \lt \{
-\c^{\dot \a}
\pa_{\a\dot\a}
\ov\c^{ \a}
-
\fr{1}{2}
F_{\dot \a\dot \b}
F^{\dot \a\dot \b}
+
D^2 \rt \}
\la{kinchi2}
\ee
The Zinn Action for the SUSY Gauge U(1) theory is
\be
 \cA_{\rm Gauge\;U(1)}
\la{chiwithzinnsout}
=\int d^4 x
\lt\{ U_{  \dot \a}
 \lt (  
i D \oC^{\dot \a}
+
F^{(\dot \a \dot \b)} 
 \oC_{\dot \b}
+ \x \cdot \pa\;
 \c^{\dot  \a}
\rt)
\rt.
\eb+ \ov U_{ \a}
 \lt (
-i   D C^{ \a}
+
 F^{( \a  \b)} 
 C_{ \b}
+ \x \cdot \pa\;
 \ov \c^{  \a}
\rt)
\ee
\be
+
  \Xi
\lt (
\fr{i}{2}
 \pa^{\a  \dot \a} 
\ov\c_{\a} \oC_{\dot\a}
-\fr{i}{2}
 \pa^{\a  \dot \a} 
 \c_{ \dot\a}  C_{\a}
+ \x \cdot \pa\;
D
\rt )
\eb+
\W_{ \a\dot \a}
\lt ( 
\pa^{ \a\dot \a}
 \w 
 +  \c^{\dot \a}
C^{\a}
+ 
 \ov\c^{\a}\oC^{\dot \a} 
+ \x \cdot \pa\;
V ^{\a\dot \a}
\rt )
\la{gaugetrans}
\ee

\be
+K
\lt ( V^{\b\dot \b}
C_{\b} \oC_{\dot \b}
+ \x \cdot \pa\;
\w
\rt )
+  
J'  
\lt (
L + \x \cdot \pa\;
\h'
\rt )
\lt.
\ebp+ 
 \D 
\lt (
C_{\b} \oC_{\dot \b}
\pa^{\b \dot \b}
\h'
+ \x \cdot \pa\;
L
\rt )
\rt \}
\ee
The Zinn Action for the SU(2) SUSY Gauge theory is very similar, but with SU(2)  rotations and an index $a$.  Again, for the Gluons that bind the Quarks, we need to also add an SU(3) SUSY gauge theory, which is also similar.

\refstepcounter{orange}
{\bf \theorange}.\;{\bf \PB:}
Here we need to assemble the various \PB s that are needed for
the various Actions. Here is the \PB\ that applies to the Action
in this paper:
\be
\cP_{\rm Total}[\cA]=
\cP_{\rm Gauge}[\cA]
+\cP_{\rm Half \;Chiral}[\cA]
+\cP_{\rm Un-Chiral}[\cA]
+\cP_{\rm SUSY}[\cA]
\la{newPB}
\ee

The \PB s $\cP_{\rm Half \;Chiral}[\cA]$ and $\cP_{\rm Un-Chiral}[\cA]
$ have been written down in \ci{four} for both the Majorana \HC\ \ci{majhalfchiral} and the Dirac \HC\ \ci{dirachalfchiral} cases, and also for the Dirac \UC\ \ci{majunchiral} case.  Here we just need to adjust them by adding the appropriate indices for the three Chiral Higgs Multiplets, which we will take to \HC\ form; and for the three flavours of Quark and Lepton Left SU(2) Doublets and Right SU(2) Singlets, which we will take to \UC\ form as discussed below in \pgh\ \ref{lepsandquarks}.

The SUSY Gauge \PB s are:
\be
\cP_{\rm Gauge}[\cA]= 
\cP_{\rm U(1)}[\cA]+
\cP_{\rm SU(2)}[\cA]+
\cP_{\rm SU(3)}[\cA]
\ee

\be\cP_{\rm U(1)}[\cA]=
\eb\int d^4 x \lt \{
\fr{\delta \cA}{ \delta U_{\dot \a}}
\fr{\delta \cA}{ \delta \chi^{\dot \a}}
+
\fr{\delta \cA}{ \delta \ov U_{  \a}}
\fr{\delta \cA}{ \delta \ov \chi^{  \a}}+
\fr{\delta \cA}{ \delta \W_{\a\dot \a}}
\fr{\delta \cA}{ \delta V^{\a\dot \a}}
+
\fr{\delta \cA}{ \delta  \X}
\fr{\delta \cA}{ \delta D }
+\fr{\delta \cA}{ \delta K}
\fr{\delta \cA}{ \delta \w}
+\fr{\delta \cA}{ \delta J' }
\fr{\delta \cA}{ \delta \h'}
\rt \}
\ee

As usual, we need to
add the Action $\cA_{\rm SUSY}= -h^{\a \dot \a}
C_{\a}\oC_{ \dot \a}$ to the Action to close the algebra.  
$\cP_{\rm SU(2)}[\cA]
$  and $\cP_{\rm SU(3)}[\cA]
$ are similar to $\cP_{\rm U(1)}[\cA]$.

\refstepcounter{orange}\la{abbrevs}
{\bf \theorange}.\;{\bf 
Covariant Derivatives and Rotations:}
Now we want to couple our  Chiral Multiplets to SUSY Gauge theory.  Here we just treat these as normal chiral Multiplets.  Later we will convert them to \HCM s.
Here we will use the abbreviations

\be
 {\cal D}^i_{ j \a\dot\a}
=
\lt (
 \pa_{ \a\dot\a}\d_j^i
+ i \fr{1}{2} g_1  V_{ \a\dot\a}\d_j^i
+ i \fr{1}{2} g_2 V^a_{ \a\dot\a}\s^{ai}_j
\rt )
\eb{\ov {\cal D}}^i_{ j \a\dot\a}
=
\lt (
 \pa_{ \a\dot\a}\d_j^i
- i \fr{1}{2} g_1  V_{ \a\dot\a}\d_j^i
- i \fr{1}{2} g_2 V^a_{ \a\dot\a}\s^{ai}_j
\rt )
\la{coderivs}\ee
and
\be
{\cal R}_{j}^i=
\lt (
- i \fr{1}{2}g_1  \w \d_j^i
-  i \fr{1}{2}g_2 
\s^{ai}_j
\w^a
\rt )
;\;{\cal {\ov R}}_{Hj}^i=
\lt ( i \fr{1}{2}g_1  \w \d_j^i
+  i \fr{1}{2} g_2 
\s^{ai}_j
\w^a
\rt )\ee

\refstepcounter{orange} \la{HLD}
{\bf \theorange}.\;{\bf 
The HL Chiral Multiplet Coupled to Gauge theory:}
This is a Higgs Multiplet with quantum numbers
$(\fr{1}{2},\fr{1}{2})$ 
of $SU(2) \times U(1)$, as it appears after the $F$ auxiliaries have been integrated out of the theory.
The notation here was used in \ci{four}.
Using the abbreviations in \pgh\ \ref{abbrevs}, we can write:
\be
\cA_{\rm HL}
= \int d^4 x \; \lt \{
-
\ov\f_{{HL}i}^{ \a}
 {\cal D}^i_{j \a\dot\a}
 \f_{{HL}}^{j\dot \a}
\ebp
+\fr{1}{ 2 }
{\cal D}^i_{k \a\dot\a} H_{L}^k \;
 {\cal {\ov D}}^{j \a\dot\a}_{i}
 \oH_{Lj} 
\la{leftfermi}
+
\G_{{HR}i}
 \lt [
 \f^i_{{HL} \dot \d}
\oC^{\dot \d}
+{\cal R}_j^i  H_{L}^j
+
\x \cdot \pa H_L^i
\rt ]
\rt.
\ee

\be
\lt.
+
 \ov\G_{HR}^i  
 \lt [
  \ov\f_{{HL}  i \d}
C^{ \d}
+
{\cal {\ov R}}_i^j \oH_{Lj}
+
\x \cdot \pa \oH_{Li}
\rt ]
\ebp+
 Z_{{HR}i}^{\dot \a}
\lt [
 {\cal D}^i_{j \a\dot\a}
H_{L}^j
C^{\a} 
+{\cal R}_j^i
\f^j_{{HL}\dot \a}
+
\x \cdot \pa \f_{H L\dot \a}^i
\rt ]
\rt.
\ee

\be
\lt.
+ 
 \oZ_{{HR}}^{ j\a}
\lt [
 {\cal {\ov D}}^i_{j \a\dot\a}
\oH_{Li}
\oC^{\dot\a} 
+
{\cal {\ov R}}_j^i \ov \f_{{HL}i\a}
+
\x \cdot \pa \ov \f_{H L j \a} 
\rt ]
\ebp
+\fr{1}{2}
\lt (
 - i g_1  \c^{\dot \a} 
\f^i_{{HL}\dot \a} 
\oH_{Li}
+
i g_1  \ov\c^{  \a} 
\ov\f_{{HL} i\a}  H^i_{L}
- g_1 D H^i_{L} \oH_{Li}
\rt )
\rt.
\ee

\be
\lt.+
\fr{1}{2}
\lt (
-
i g_2 \c^{a\dot \a} \s^{ai}_{j}
\f^j_{{HL}\dot \a} \oH_{Li}
\rt.
\ebp
\lt.
+ 
i g_2 \ov\c^{ a \a} \s^{ai}_{j}
\ov\f_{{HL} i\a}  H^j_{L}
- g_2 D^a \s^{ai}_{j}   H^j_{L}
\oH_{Li}
\rt)
\rt \}
\ee

\refstepcounter{orange}\la{HRD}
{\bf \theorange}.\;{\bf The HR \CM\ Coupled to Gauge theory:}
This is a Higgs Multiplet\footnote{The $\fr{1}{2}$ representation of SU(2) is pseudoreal so we can take $(\fr{-1}{2},\fr{-1}{2})$ or $(\fr{1}{2},\fr{-1}{2})$ here.  We have chosen the former.} with quantum numbers
 $(\fr{-1}{2},\fr{-1}{2})$
of $SU(2) \times U(1)$, as it appears after the $F$ auxiliaries have been integrated out of the theory.
The notation here was used in \ci{four}.
Using the abbreviations in \pgh\ \ref{abbrevs}, we can write:
\be
\cA_{\rm HR}
= \int d^4 x \; \lt \{
-
\ov\f_{{HR}}^{j \a}
 {\cal {\ov D}}^i_{j \a\dot\a}
\f_{{HR}i}^{\dot \a}
+\fr{1}{ 2 }
 {\cal {\ov D}}^i_{j \a\dot\a}
  H_{Ri}
 {\cal  D}^{j \a\dot\a}_{k }
 \oH_{R}^{k}
\la{rightfermi}\ebp
+
 \ov\G_{HLi}  
 \lt [
  \ov\f^i_{{HR} \d}
C^{ \d}
+
{\cal R}_j^i
\oH_{R}^j
+
\x \cdot \pa \oH_{R}^i
\rt ]
+
\G_{{HL}}^i
 \lt [
 \f_{{HRi} \dot \d}
\oC^{\dot \d}
+
{\cal {\ov R}}_i^j H_{Rj}
+
\x \cdot \pa H_{Ri}
\rt ]
\ebp
+ 
 \oZ_{{HL}i}^{ \a}
\lt [
 {\cal D}^i_{j \a\dot\a}
 \oH_{R}^j
 \oC^{\dot \a} 
+{\cal R}_j^i  {\ov \f}^j_{{HR}\a}
+
\x \cdot \pa \ov \f^i_{{HR}\a}
\rt ]
\ebp
+
 Z_{{HL}}^{j\dot \a}
\lt [
 {\cal {\ov D}}^i_{j \a\dot\a}
 H_{Ri}
  C^{\a} 
+ {\cal {\ov R}}_j^i 
\f_{{HR}i\dot \a}
+
\x \cdot \pa \f_{{HR}j\dot \a}\rt ]
\la{ZHLterm}
\ebp
+
\fr{1}{2}
 \lt (
+ 
i g_1 \c^{\dot \a} 
\f_{{HRi}\dot \a} \oH_{R}^i
-i g_1 \ov\c^{  \a} 
\ov\f^i_{{HR}\a}  H_{Ri}
+  g_1 D \oH_{R}^i H_{Ri} 
\rt) 
\ebp
+ 
\fr{1}{2}
\lt (+
i g_2 \c^{a\dot \a} 
\ov\s^{ai}_{\;\;\;j}
\f_{{HRi}\dot \a}  \oH_{R}^j
-
i g_2 \ov\c^{ a \a}
\ov\s^{ai}_{\;\;\;j}\ov\f^j_{{HR} \a}  H_{Ri} 
+  g_2 D^a \ov\s^{ai}_{\;\;\;j}
 \oH_{R}^j H_{Ri}  
\rt \}
\rt \}
\ee

\refstepcounter{orange}\la{GD}
{\bf \theorange}.\;{\bf The G \CM\ 
Coupled to SU(2):}
This is a Higgs Multiplet with quantum numbers
$(1,0)$ 
of $SU(2) \times U(1)$, as it appears after the $F$ auxiliaries have been integrated out of the   theory. 

\be
{\cal A}_{\rm G}
=  \int d^4 x \; \lt \{
-\f^{a \dot \a}
\lt (
\pa_{\a\dot\a}\d^{ac}
+ g_2 \ve^{abc}V^b_{\a\dot\a}
\rt )
\ov\f^{c\a}
\ebp
+ \fr{1}{2}
\lt (
\pa_{\a\dot\a}\d^{ac}
+ g_2 \ve^{abc}V^b_{\a\dot\a}
\rt )
S^c \lt (
\pa^{\a\dot\a}\d^{ae}
+ g_2 \ve^{ade}V^{d \a\dot\a}
\rt )
 \oS^e 
\la{GScalarterm}
\ebp+
\oG^a
\lt (
\ov\f^a_{\d}
C^{ \d}
+ g_2 \ve^{ade} \oS^d \w^e
+ \xp  \oS^a\rt )
\ebp
+
\G^a
\lt(
\f^a_{\dot\d}
\oC^{\dot \d}
+ g_2 \ve^{ade} S^d \w^e
+ \xp S^a  \rt )
\ebp+
  Z^a_{\dot \a} 
\lt (
  D^{ab\;\a \dot \a} S^b
C_{\a}
+ g_2 \ve^{ade} \f^{d\dot \a} \w^e
+ \xp  \f^{a\dot \a} \rt )
\ebp
+ \ov Z^a_{ \a} 
\lt (
  D^{ab\;\a \dot \a} \oS^b
\oC_{\dot\a}  
+ g_2 \ve^{ade} \ov \f^{d \a} \w^e
+ \xp \ov \f^{a\a}  \rt )
\ebp
-
 g_2 \ve^{abc} 
\c^{a\dot \a} \f^b_{\dot \a} \oS^c
+
 g_2 \ve^{abc} 
\ov \c^{a  \a} \ov \f^b_{  \a} S^c
- g_2 \ve^{abc} D^a S^b
\oS^c
\rt\}
\ee

\refstepcounter{orange}\la{Tterms} 
{\bf \theorange}.\;{\bf Yukawa Terms and
Mass Terms from the Superpotential:}
With these \CM s we can construct only the following two mass terms and one Yukawa type coupling, assuming that the Action has the renormalizable dimensions. 
We are writing these as they appear after the $F$ auxiliaries have been integrated out of the   theory:
\be 
\cY_{3}=
\fr{1}{2}M_{3}  \int d^4 x \lt \{
\f^{a\dot \a}
\f^a_{\dot \a}
\rt \}
\ee
\be
\cY_{4}=
M_4 \int d^4 x \lt \{
\f_{HL}^{i\dot \a}
\f_{{HRi}\dot \a}
\rt \}
\ee
\be
\cY_{5}=
  \sqrt{2} i g_{5}
\int d^4 x \lt \{
\s^{ai}_{j}
S^a 
\f_{HL}^{j\dot \a}
\f_{{HRi}\dot \a}
+ 
\s^{ai}_{j}
H_{L}^j
\f_{HRi}^{\dot \a}
\f^a_{\dot \a}
+\s^{ai}_{j}
H_{Ri} 
\f_{HL}^{j\dot \a}
\f^a_{\dot \a}
\rt \}
\ee

These masses $M_3, M_4$ and the dimensionless coupling $g_5$ can be chosen to be real.

\refstepcounter{orange}\la{CTs}
{\bf \theorange}.\;{\bf \CT s
from the superpotential:}
The terms here can be found by including $F$ auxiliaries and then integrating them out.
This result is another application of theorem 1 of \pgh\ 8 in \ci{four} relating to the integration of auxiliaries in the presence of the \PB.

\be
\cA_{\rm CT, L}
=
- \int d^4 x 
\lt |
 Z^i_{{HL}   \dot\a}
  \oC^{\dot\a}+
M_{4}
H_{L}^{i} 
+
\sqrt{2}
i g_{5}
\s^{ai}_{j}
H_{L}^{j} 
S^a
\rt |^2
\ee

\be
\cA_{\rm CT,R}
=
- \int d^4 x 
\lt |
 Z_{{HRi}  \dot \a}
  \oC^{\dot\a}+
M_{4}
H_{Ri} 
+\sqrt{2}
i g_{5}
\s^{aj}_{i}
H_{Rj}
S^a
\rt |^2
\ee

\be
\cA_{\rm CT,G}
=
-\int d^4 x 
\lt |
 Z^a_{\dot \a} \oC^{\dot \a}
+
 M_{3} 
S^a 
+ 
\sqrt{2}
i g_{5}
\s^{ai}_{j}
H_L^j 
H_{Ri} 
\rt  |^2
\ee

\refstepcounter{orange}
{\bf \theorange}.\;{\bf Pure Field
Yukawa Terms:}
The \CTR s that implement \HC\ theories here are described and explained in \ci{four}. 
There is one Dirac Multiplet and one Majorana Multiplet here in terms of the language used there. Using the \CTR s in \ci{fourgenfordirac} for $H$, and
\ci{fourgenformajorana} for $S$, and dropping all Zinn Sources, we make the  substitutions:
\be
H_L^i 
\ra -\;\fr{1}{\sqrt{2}}
 H^i
;\;H_{Ri} 
\ra 
 \fr{1}{\sqrt{2}}
\oH_i
;\;S^a
\ra - i \fr{1}{\sqrt{2}} G^a
\ee
These yield 
\be
\cA_{\rm CT\;L}
=
-\fr{1}{2}\int d^4 x 
\lt |
M_{4}
H^{i} 
+
 g_{5}
\s^{ai}_{j}
H^{j} 
G^a
\rt |^2
\ee

\be
\cA_{\rm CT\;R}
=
-\fr{1}{2}\int d^4 x 
\lt |
M_{4}
\oH_{i}
+
 g_{5}
\s^{aj}_{i}
\oH_{i}
G^a
\rt
|^2
\la{LandRfterms}
\ee
\be
\cA_{\rm CT\;G}
=
-\fr{1}{2}\int d^4 x 
\lt |
  M_{3} 
G^a 
+
 g_{5}
\s^{ai}_{j}
H^j 
\oH_{i} 
\rt  |^2
\la{GFterms}
\ee
The sum of these, with a minus sign, is the \EP\ for this theory\footnote{
The terms
\be
- g_1 D   H^i    \oH_i   
- g_2 D^a \s^{ai}_{j}   H^j    \oH_i    
\ee
do not appear in the effective potential before integration of
the auxiliaries, because they cancel between the Actions
$\cA_{\rm HR\;\CHC}$
 and 
$\cA_{\rm HL\;\CHC}$
 when we pick out the pure field parts.  Similarly, the term $\ve^{abc} D^a G^bG^c=0$ also does not contribute.}.

\refstepcounter{orange}\la{VEVS}
{\bf \theorange}.\;{\bf  VEVs:}
This \EP\ is positive semidefinite.  It has two different zero energy configurations. 
One has zero VEVs and it does not spontaneously break the gauge group $SU(2) \times U(1)$.  The other has non-zero VEVs, and it implements \GSBS.
For that configuration we have:

\be
\lt < G^a \rt >
=M_G \d^{a3} 
;\;
\lt < H^i \rt >
=M_H \d^{i}_1 
\la{HVEV}
\ee
To get zero energy we require
 \be
M_G
=\fr{-M_{4}}{ g_{5}}
;\;M_H^2
=-\fr{M_G  M_{3}}{ g_{5}}
=\fr{M_{3}  M_{4}}{ g_{5}^2}
\la{hvalue}
\ee

\refstepcounter{orange} \la{shiftsdefe}
{\bf \theorange}.\;{\bf The Pure Field Part of the \HC\ 
Action and the Spectrum of Masses:}
 It is straightforward  to implement the shift and canonical transformation from the Chiral form to the physical \HC\ form of this Action. To get the mass spectrum we only need a small part of this Action. Starting with the Action above in \pgh s  \ref{HLD},\ref{HRD},\ref{GD},\ref{Tterms} and \ref{CTs}, and using \ci{fourgenfordirac} for $H$, and
\ci{fourgenformajorana} for $S$,  and the VEV s in equation
 (\ref{HVEV}), and dropping all Zinn Souces,
we substitute the following
\be
H_L^i\ra 
\fr{1}{\sqrt{2}}\lt (  - H^i -M_H \d^i_1  \rt)
;\;H_{Ri}
\ra
 \fr{1}{\sqrt{2}}\lt ( \oH_i  +M_H \d_i^1  \rt) 
;\;S^a
\ra 
\fr{1}{\sqrt{2}}
\lt (  - i G^a - i M_G \d^{a3} \rt )
\la{subtogetpure}
\ee
To find the spectrum we only need the terms that involve the fields, and we only need the terms which are quadratic in the fields.
  So we    drop  all Zinn Source terms and also set all terms with the Antighost $\h$ to zero, because they are not quadratic.

\refstepcounter{orange}\la{neutghostsec}
{\bf \theorange}.\;{\bf The Neutral Goldstone  Boson and its Ghost Action:}

We define two real fields, which are the (properly renormalized\footnote{ From (\ref{lefthfield}) and the similar term from (\ref{rightfermi}), we get the kinetic terms  $ \fr{1}{2} \int d^4 x \pa_{\a \dot \a} H^1 \pa^{\a \dot \a} \oH_1 =\int d^4 xH^1 \Box \oH_1  = \fr{1}{2}\int d^4 x\lt (  H_{\rm Im 1}\Box H_{\rm Im 1} + H_{\rm Re 1}\Box H_{\rm Re 1} \rt )$})  real and imaginary parts of the complex field $H^1$:
\be
H_{\rm Im 1} = 
\lt   [
\fr{-iH^1   +i\oH_1 }{\sqrt{2}}
\rt ] ; \;H_{\rm Re 1} = 
\lt   [
\fr{H^1  + \oH_1 }{\sqrt{2}}
\rt ] 
\ee

After the substitution in (\ref{subtogetpure}), the Action in equations (\ref{leftfermi}) with the covariant derivatives in \pgh\ \ref{abbrevs} yields  
\be
+\fr{1}{ 2 }
\lt (
 \pa_{ \a\dot\a}\d_j^i
+ i \fr{1}{2} g_1  V_{ \a\dot\a}\d_j^i
+ i \fr{1}{2} g_2 V^a_{ \a\dot\a}\s^{ai}_j
\rt )
\fr{1}{\sqrt{2}}\lt (  - H^j -M_H \d^j_1  \rt)
\eb
\lt (
 \pa_{ \a\dot\a}\d_i^k
- i \fr{1}{2} g_1  V_{ \a\dot\a}\d_i^k
- i \fr{1}{2} g_2 V^a_{ \a\dot\a}\s^{ak}_i
\rt )
\fr{1}{\sqrt{2}}\lt (  - \oH_k -M_H \d^1_k  \rt)
\la{lefthfield}\ee
and there is a similar term from 
Equation (\ref{rightfermi}).
These two terms from Equations (\ref{leftfermi}) and (\ref{rightfermi})  
 contain the `neutral cross terms'
\be
- \fr{1}{2 \sqrt{2} } M_H 
  g_1
H_{\rm Im 1} \pa_{ \a\dot\a}
V^{ \a\dot\a}     
- \fr{1}{2 \sqrt{2} } M_H 
  g_2
H_{\rm Im 1} \pa_{ \a\dot\a}
V^{3 \a\dot\a}     
\la{neutcrossfromact}
\ee
The standard procedure is to eliminate these cross terms using the `t Hooft trick \ci{taylor}.  One chooses  a Faddev Popov Ghost and Gauge Fixing Action 
so that there is a term that can cancel the cross terms.  Here, for the first term in
(\ref{neutcrossfromact}), we choose
 \be
\cA_{\rm GGF}= 
\int d^4 x \d
 \lt \{
 \h 
 \lt ( p_{0} L -  \fr{2 p_{0} }{2 \sqrt{2} } M_H
  g_1
H_{\rm Im 1}
 +    \pa_{ \a\dot\a}
V^{ \a\dot\a}     
 \rt )
 \rt \}
\eb=
\int d^4 x 
 \lt \{
 L 
 \lt ( p_{0} L -  \fr{ 2 p_{0} }{2 \sqrt{2} } M_H
  g_1
H_{\rm Im 1}
 +    \pa_{ \a\dot\a}
V^{ \a\dot\a}     
 \rt )
 \rt \}+ \cdots
\la{fromLvar} \ee
where $p_{0}$ is a dimensionless gauge parameter.  In the above, we used
\be
\d \h = L
\ee

Now we  complete the square for the $L$ terms, and write (\ref{fromLvar}) in the form:
\be
\cA_{\rm GGF}
= 
\eb\int d^4 x 
 \lt \{
  p_{0}   
 \lt (   L - \fr{1}{2p_{0} }\lt [
 \fr{2 p_{0}}{2 \sqrt{2}} M_H
  g_1
H_{\rm Im 1}
 -    \pa_{ \a\dot\a}
V^{ \a\dot\a} \rt]     
 \rt )^2
 \rt \}
\eb
-\int d^4 x 
 \lt \{
  p_{0}   
 \lt (    \fr{1}{2p_{0} }\lt [
 \fr{ 2p_{0}}{2 \sqrt{2}} M_H
  g_1
H_{\rm Im 1}
 -   \pa_{ \a\dot\a}
V^{ \a\dot\a} \rt]     
 \rt )^2
 \rt \}
\ee
  Shifting and integrating the auxiliary $L$ yields

\be
-\int d^4 x 
 \lt \{
 \fr{1}{4p_{0} } 
 \lt (    \lt [
 \fr{2 p_{0} }{2 \sqrt{2}} M_H
  g_1
H_{\rm Im 1}
 -   \pa_{ \a\dot\a}
V^{ \a\dot\a} \rt]     
 \rt )^2
 \rt \}
 \la{neutcrossfromactrem}\ee
The cross term here  cancels the term 
(\ref{neutcrossfromact})
above, leaving us with

 \be
\cA_{\rm GGF}
= - \fr{1}{4p_{0} } \int d^4 x 
\eb \lt \{
 \lt ( 
\lt [
 \fr{2 p_{0} }{2 \sqrt{2} }M_H 
  g_1
H_{\rm Im 1}
\rt ]^2 
+  \lt [
   \pa_{ \a\dot\a}
V^{ \a\dot\a}     
\rt ]^2 \rt ) 
\rt \}
\la{aftercrossing}
\ee

Now we want to choose the Feynman gauge for simplicity.
The kinetic terms for the  vector Bosons are
\be
\cA_{\rm Kinetic \;\c}
=
 \int d^4 x \; \lt \{
-
\fr{1}{2}
F_{\dot \a\dot \b}
F^{\dot \a\dot \b}
-
\fr{1}{2}
F^a_{\dot \a\dot \b}
F^{a\dot \a\dot \b}
\rt\}
\eb
=\fr{-1}{2}\int d^4 x 
\lt \{
 V^{\a \dot \a}
\lt (
\Box
V_{\a \dot \a} 
+
\fr{1}{2}
\pa_{\a \dot \a} \pa_{\g \dot \d} V^{\g  \dot \d} 
\rt )
\ebp+ V^{a\a \dot \a}
\lt (
\Box
V^a_{\a \dot \a} +
\fr{1}{2}
\pa_{\a \dot \a} \pa_{\g \dot \d} V^{a\g  \dot \d} 
\rt )
\rt \}
\la{normofvectors}
\ee
So we have in this sector
\be
\cA_{\rm Kinetic}
+\cA_{\rm GGF}
= \int d^4 x 
 \lt \{
- \fr{1}{4p_{0} } \lt (   \lt [
   \pa_{ \a\dot\a}
V^{ \a\dot\a}     
\rt ]^2 \rt ) 
 \rt \}
\eb
+\int d^4 x 
\lt (
\fr{-1}{2} V^{\a \dot \a}
\Box
V_{\a \dot \a} 
+
\fr{1}{4}
\pa_{\a \dot \a} V^{\a \dot \a}
 \pa_{\g \dot \d} V^{\g  \dot \d} 
\rt )
 \ee
and so if we choose
\be
p_{0} =1
\ee
then we get the Feynman gauge, where the equation of motion does not have a 
$\pa_{\a \dot \a}
 \pa_{\g \dot \d} V^{\g  \dot \d} $ term.

This leads from (\ref{aftercrossing})
to the mass term 
 \be
\cA_{\rm Neutral\;
Goldstone\; Mass\;0}
=- \fr{1}{8}   M_H^2   g_1^2
 \int d^4 x 
 H_{\rm Im 1}^2 
\la{zeromass} \ee

 Then, for the second term in
(\ref{neutcrossfromact}),  with the similar technique we get
 \be
\cA_{\rm Neutral\;
Goldstone\; Mass\;3}
=- \fr{1}{8}  M_H^2   g_2^2
 \int d^4 x 
 H_{\rm Im 1}^2 
\la{threemass}
 \ee
and the sum of these two in ({\ref{zeromass})  and ({\ref{threemass}) is
the total neutral Goldstone mass term for this gauge choice:
 \be
\cA_{\rm  \rm Neutral\;Goldstone\; Mass}
= \int d^4 x 
 \lt \{
 - \fr{1}{2} 
M_H^2
 \lt (
 g_1^2
+ g_2^2
\rt )
H_{\rm Im 1}^2  
\rt \}
\eb
\;{\rm where}\;
 M_Z^2=
 \fr{1}{ 4}
M_H^2\lt (g_1^2 +g_2^2\rt )
\ee With these normalizations we get 
\be
\lt ( \Box  - M_Z^2 \rt )H_{\rm Im 1}=0
\ee
So this  is the Goldstone Boson mass for the $ H_{\rm Im 1}$ in the Feynman gauge.

\refstepcounter{orange}
{\bf \theorange}.\;{\bf  Neutral Vector Boson:}
Now that we have cancelled the above mixing term in equation (\ref{neutcrossfromact}) using the 't Hooft cross term in (\ref{neutcrossfromactrem}), we can look at the remaining  Neutral Vector Boson Mass Terms from  the substitution in (\ref{subtogetpure}), the Action in equations (\ref{leftfermi}) and (\ref{rightfermi}), with  the covariant derivatives in \pgh\ \ref{abbrevs} yields: 
\be
\int d^4 x
 \fr{1}{ 8 }
M_H^2
\lt \{ g_1
 V_{ \a\dot\a} 
 +g_2   V^3_{ \a\dot\a}
\rt \}
\lt \{ g_1
 V^{ \a\dot\a} 
 +g_2   V^{3  \a\dot\a}
\rt \}
\ee
So we define the new normalized vector Boson
\be
Z_{\a \dot \a}= \lt \{ \sin \q_W
 V_{ \a\dot\a} 
 - \cos \q_W V^3_{ \a\dot\a}
\rt \}
\la{defZ}
\ee
where
\be
 \sin \q_W= \fr{g_1}{\sqrt{g_1^2 +g_2^2}}
;\; \cos \q_W= \fr{g_2}{\sqrt{g_1^2 +g_2^2}}
;\; \tan \q_W = \fr{g_1}{g_2}
\ee
Then the above is\footnote{Comparison with \ci{PDGstandardModel}
 shows that
$
g_1= -g'
;\;g_2= g
$}

\be
\int d^4 x
 \fr{1}{ 2} M_Z^2
Z_{ \a\dot\a} 
Z^{ \a\dot\a} 
\;{\rm where}
\; M_Z^2=
 \fr{1}{ 4}
M_H^2\lt (g_1^2 +g_2^2\rt )
\ee
Then using the term (\ref{normofvectors}), we see that we get the equation of motion
\be
\lt (
\Box - M_Z^2\rt )
Z_{\a \dot \a}=0
\ee
which confirms that  this is indeed the mass of the Z Boson.

\refstepcounter{orange}
{\bf \theorange}.\;{\bf   The Photon:}
  We define the photon as the normalized vector Boson orthogonal to the $Z$ in equation (\ref{defZ}) above.  It has the form:
\be
A_{\a \dot \a}= \lt \{ \cos \q_W
 V_{ \a\dot\a} 
 + \sin \q_W V^3_{ \a\dot\a}
\rt \}
\la{defA}
\ee
There are no remaining mixing or mass terms for this orthogonal combination, so the 
photon is massless.  In order to make contact with experiment we note that the coupling of this vector Boson to a charged field can be found  from  the substitution in (\ref{subtogetpure}), the Action in equations (\ref{leftfermi}) and (\ref{rightfermi}),  and the covariant derivatives in (\ref{coderivs}).  This yields the value of the electromagnetic coupling constant $e$:
\be
e=g_2  \sin \q_W
\ee

\refstepcounter{orange}
{\bf \theorange}.\;{\bf   The Two Neutral Higgs Bosons:}
We still have some neutral terms left, and they are independent of the terms that appear in the Goldstone Action above for the Z.
  Here we will lower the index to prevent confusion.
\be
G^3\ra G_3
\ee
The Action here comes from the expansion of (\ref{LandRfterms}) and 
(\ref{GFterms}) using the substitution in (\ref{subtogetpure}).  It is

\be
\cA_{\rm Neutral \;Scalars}
=
 \int d^4 x \lt \{
g_{5}^2
M_H^2     (G_3)^2
\rt \}
+ \fr{1}{2}  \int d^4 x  \lt \{
 M_{3} 
   G_3  
+ 
g_{5}  M_H
\sqrt{2} H_{\rm Re 1}  
\rt  \}^2
\ee

Noting the Scalar Kinetic terms  in equations (\ref{leftfermi}) and (\ref{rightfermi}), we see that the
 equations of motion are

\be
\Box 
\lt ( \ba{c} G_3 \\ H_{\rm Re 1} \\ \ea
\right)
 =
 \left( \ba{cc} 2\lt (
g_{5}^2
M_H^2  + \fr{1}{2}
 M_{3}^2\rt )  & g_{5} \sqrt{2} M_H  M_{3} 
  \\ g_{5} \sqrt{2} M_H  M_{3} 
 &2 g_{5}^2  M_H^2
\\ \ea
\right)
\lt ( \ba{c} G_3 \\ H_{\rm Re 1} \\ \ea
\right) \ee
 
The eigenvalues of this mass matrix are the squares of the two Neutral  Higgs masses and so we get:
\be
M_{\rm HO \;Light}=
\sqrt{
\frac{1}{2} \left(4
   g_5^2
   M_H^2+M_3^2-\sqrt{M_3^4+8
   g_5^2 M_H^2 
   M_3^2}\right)
}\ee
\be
M_{\rm HO \;Heavy}=
\sqrt{\frac{1}{2} \left(4
   g_5^2
   M_H^2+M_3^2+\sqrt{M_3^4+8
   g_5^2 M_H^2 
   M_3^2}\right)}
\ee

\refstepcounter{orange}\la{neutfermoion}
{\bf \theorange}.\;{\bf  The Neutral Fermions:} 
There are 
five neutral Majorana Fermions in this Gauge/Higgs sector.  The squared masses  are the eigenvalues of the product of the matrix
\be
\lt (
\ba{c|ccccc}
&\c&\c_3&\f_L&\f_R&\f_3\\
\hline
\c &&&i M_{1}&i M_{1}\\
\c_3&&&i M_{2}&i M_{2}&\\
\f_L &i M_{1}&i M_{2}&&&i 
  M_{5}\\
\f_R & i M_{1}&i M_{2}&&&-i 
  M_{5}\\
\f_3 &&&i 
  M_{5}&-i  M_{5}&M_3\\
\ea
\rt )
\la{neutmassmat}\ee

with its \CC. This matrix is
\be
\left(
\begin{array}{ccccc}
 2 M_1^2 & 2 M_1 M_2 & 0 & 0 & 0 \\
 2 M_1 M_2 & 2 M_2^2 & 0 & 0 & 0 \\
 0 & 0 & M_1^2+M_2^2+M_5^2 & M_1^2+M_2^2-M_5^2 & i M_3 M_5 \\
 0 & 0 & M_1^2+M_2^2-M_5^2 & M_1^2+M_2^2+M_5^2 & -i M_3 M_5 \\
 0 & 0 & -i M_3 M_5 & i M_3 M_5 & M_3^2+2 M_5^2 \\
\end{array}
\right)
\la{squareneut}
\ee

Here we have defined:
\be
M_1= \fr{g_1}{2 \sqrt{2}} M_H
;\;M_2= \fr{g_2}{2 \sqrt{2}} M_H
;\; M_5= g_5  M_H
\ee

 The five eigenvalues of the matrix  (\ref{squareneut}) 
 are:
\be
\left\{0,M_Z^2,M_Z^2,M_{\rm HO \;Light}^2, M_{\rm HO \;Heavy}^2
\right\}
\la{neutmasseig}
\ee
So the masses of two of these Majorana Fermions are the same as $M_Z^2$, two more match the masses of the two neutral Higgs Bosons, and there is a zero mass Majorana Fermion.

\refstepcounter{orange}\la {startcharged}
{\bf \theorange}.\;{\bf The Charged Goldstone  Boson:}
After the substitution in (\ref{subtogetpure}), the Action in equations (\ref{leftfermi})  with 
the covariant derivatives in \pgh\ \ref{abbrevs} yields: 
\be
+\fr{1}{ 2 }
\lt (
 \pa_{ \a\dot\a}\d_j^i
+ i \fr{1}{2} g_1  V_{ \a\dot\a}\d_j^i
+ i \fr{1}{2} g_2 V^a_{ \a\dot\a}\s^{ai}_j
\rt )
\fr{1}{\sqrt{2}}\lt (  - H^j -M_H \d^j_1  \rt)
\eb
\lt (
 \pa_{ \a\dot\a}\d_i^k
- i \fr{1}{2} g_1  V_{ \a\dot\a}\d_i^k
- i \fr{1}{2} g_2 V^a_{ \a\dot\a}\s^{ak}_i
\rt )
\fr{1}{\sqrt{2}}\lt (  - \oH_k -M_H \d^1_k  \rt)
\ee
and there is a similar term from (\ref{rightfermi}).
In addition we have a term  from (\ref{GScalarterm}):

\be
+ \fr{1}{2}
\lt (
\pa_{\a\dot\a}\d^{ac}
+ g_2 \ve^{abc}V^b_{\a\dot\a}
\rt )
\fr{1}{\sqrt{2}}
\lt (  - i G^c - i M_G \d^{c3} \rt )
\eb
\lt (
\pa^{\a\dot\a}\d^{ae}
+ g_2 \ve^{ade}V^{d \a\dot\a}
\rt )
\fr{1}{\sqrt{2}}
\lt (   i G^e + i M_G \d^{e3} \rt )
\ee

These three terms from (\ref{leftfermi}), (\ref{rightfermi}) and  (\ref{GScalarterm}) 
 contain the `charged cross terms'
\be
{\rm Cross\;Terms}=
+   g_2  
\lt [ M_H 
\fr{1}{2\sqrt{2}}  H_{\rm Re 2}
-  M_G 
  \fr{1}{2}
  G_1 \rt ]
     \pa_{ \a\dot\a}  V^{2  \a\dot\a}
\la{onechg1}
\ee

\be
- 
 g_2  
\lt [ M_H
\fr{1}{2\sqrt{2}}H_{\rm Im 2}
+ \fr{1}{2}  M_G
  G_2 \rt ]
     \pa_{ \a\dot\a}  V^{1  \a\dot\a}
\la{secchg1}
\ee
where we define two real fields, which are the (renormalized) real and imaginary parts of the complex field $H^2$:
\be
H_{\rm Im 2} = 
\lt   [
\fr{-iH^2   +i\oH_2 }{\sqrt{2}}
\rt ]; \;H_{\rm Re 2} = 
\lt   [
\fr{H^2  + \oH_2 }{\sqrt{2}}
\rt ] 
\ee
To eliminate the term in 
(\ref{secchg1}) we 
follow the procedure used above in section \ref{neutghostsec}, to eliminate these cross terms using the `t Hooft trick \ci{taylor}.  One chooses  a Faddev Popov Ghost and Gauge Fixing Action 
so that there is a term that can cancel the cross terms.  Here, we take
 \be
\cA_{\rm GGF}= \int d^4 x \d
 \lt \{
 \h^1 \ebp
 \lt (   p_{1}  L^1 
- 2 p_{1} 
   g_2 
\lt [ 
\fr{1}{2\sqrt{2}}M_H H_{\rm Im 2}
+  \fr{1}{2} M_G 
  G_2 \rt ]
  +    \pa_{ \a\dot\a}
V^{1 \a\dot\a}     
 \rt )
 \rt \}
 \ee
where $p_{1}$  is a dimensionless gauge parameter, 
To get  $\cA_{\rm Gauge\;Fixing}$ we take the term coming from
\be
\d \h^1 = L^1
\ee

and this yields, after completing the square:
 \be
\cA_{\rm Gauge\;Fixing}
= \int d^4 x 
\eb 
 p_{1} \lt \{ L^1 
- 
 \fr{1}{2 p_{1}} 
\lt [
2 p_{1}g_2\lt ( 
\fr{1}{2\sqrt{2}}M_H H_{\rm Im 2}
+  \fr{1}{2} M_G 
  G_2 \rt )
  +    \pa_{ \a\dot\a}
V^{1 \a\dot\a}     
\rt ] \rt \}^2
\eb
-
 \int d^4 x 
\eb 
\fr{1}{4 p_{1}}  \lt \{ 
\lt [ 2 p_{1}
g_2 \lt ( 
\fr{1}{2\sqrt{2}}M_H H_{\rm Im 2}
+  \fr{1}{2} M_g 
  G_2 \rt )
  +   \pa_{ \a\dot\a}
V^{1 \a\dot\a}     
\rt ] \rt \}^2
 \ee

This reduces  after a shift and integration of $L^1$.

\be
-
 \int d^4 x 
\fr{1}{4 p_{1}}  \lt \{ 
\lt [ 2 p_{1}
g_2 \lt ( 
\fr{1}{2\sqrt{2}}M_H H_{\rm Im 2}
+  \fr{1}{2} M_g 
  G_2 \rt )
  +   \pa_{ \a\dot\a}
V^{1 \a\dot\a}     
\rt ] \rt \}^2
\la{remeianfaf}
 \ee
The cross term in  (\ref{remeianfaf})    cancels the term 
in equation (\ref{secchg1}). 
Again we choose the gauge parameter so that the vector Boson is in the Feynman gauge, which means, using equation (\ref{normofvectors}), that 
\be
p_{1} = 1
\ee

Then from (\ref{remeianfaf})
 we get the mass term: 
\be
-
 \int d^4 x 
\fr{1}{8} 
\lt [ 
g_2 \lt ( 
M_H H_{\rm Im 2}
+ \sqrt{2}M_G 
  G_2 \rt )
\rt ]^2
 \ee
which generates the mixed equations of motion:

\be
\Box \lt ( \ba{c} H_{\rm Im 2} \\ G_2 \\ \ea
\right)
 =
\fr{g_2 ^2}{ 8 }   \left( \ba{cc}  M_H^2  & \sqrt{2} M_G M_H  \\ \sqrt{2}
M_G M_H   &2 M_G^2\\ \ea
\right)
 \left( \ba{c} H_{\rm Im 2} \\ G_2 \\ \ea
\rt ) \ee

 The non-zero eigenvalue here  is:
 \be
 \fr{1}{ 8} \lt (  g_2 \rt )^2 
 \left(M_H^2+2 M_G^2\right)
\ee
There is also a  zero eigenvalue here that will be  removed  by  the charged Higgs Boson, considered  below  in \pgh\ \ref{chrargedhiggs}.
Define the normalized Goldstone Boson here as
\be
G_W= \fr{1}{\sqrt{2}}\lt (  H_{\rm Im 2} +G_2\rt ) 
\ee
Then the \eqm\ comes from:
\be
\cA= 
\fr{1}{2} \int d^4 x G_W \lt ( \Box - M_W^2 \rt ) G_W
\ee
where
\be
M_W^2=
\fr{ g_2^2}{ 4 }
\lt ( M_H^2
+ 2 M_G^2  
\rt )
\ee

The term 
(\ref{onechg1}) works the same way.

\refstepcounter{orange}
{\bf \theorange}.\;{\bf The Charged  Vector Boson:} 
Now we get \be
\lt
(\fr{1}{ 8 }
 g_2^2 M_H^2
 +\fr{1}{ 4 } M_G^2 g_2^2  
 \rt )
\lt (
 V^1_{ \a\dot\a}
 V^{1  \a\dot\a}\;
+
 V^2_{ \a\dot\a}
 V^{2  \a\dot\a}\;
\rt )
]\ee 

and so we can write the Kinetic terms for this part as 

\be
\cA= 
\fr{1}{2} \int d^4 x \lt \{
V^1 \lt ( - \Box + M_W^2 \rt ) V^1
+V^2 \lt ( - \Box + M_W^2 \rt ) V^2
\rt \}
\ee
where
\be
M_W^2=
\fr{ g_2^2}{ 4 }
\lt ( M_H^2
+ 2 M_G^2  
\rt )
\ee

\refstepcounter{orange}\la{chrargedhiggs}
{\bf \theorange}.\;{\bf The Charged  Higgs Boson:} 
This  is:
\be     +       \lt (
2  M_{4}^2  H_{\rm Im 2}^2 - 2 
\sqrt{2}  M_{4}  g_{5} M_H  G_2  H_{\rm Im 2}   +
 g_{5}^2 M_H^2
   G_2^2  \rt )
\la{firstchargehiggs}
\ee
\be
+
\fr{1}{2}\lt \{
 2 
 g_{5}^2 M_H^2 H_{\rm Im 2}^2 
 -
2 \sqrt{2}  M_{3} 
 g_{5} M_H
G_2   H_{\rm Im 2} 
+ M_{3}^2 
   G_2^2  
\rt  \}
\la{secchargehiggs}
\ee

These yield
\be
( M_H^+)^2=
  \left(M_3+2
   M_4\right)^2 
   \ee

There is another similar term with the same mass corresponding to the fact that this particle is charged.

\refstepcounter{orange}\la{endcharged}
{\bf \theorange}.\;{\bf  The Charged Fermions:} 
The charged  Fermion mass terms arise after the substitutions  in the Action above.  Collecting these together yields the  matrix:
\be
\left(
\begin{array}{c|cccccc}
&\c_1&\c_2&\f_L^2&\f_R^2&\f^1&\f^2\\
\hline
\c_1& 0 & 0 & i M_{\text{2}} & i M_{\text{2}} & 0 & -i M_{\text{G2}} \\
\c_2&  0 & 0 & -M_{\text{2}} & M_{\text{2}} & iM_{\text{G2}} & 0 \\
\f_L^2&  i M_{\text{2}} & -M_{\text{2}} & 0 & 2 M_4 & i M_{\text{5}} & -M_{\text{5}} \\
\f_R^2& i M_{\text{2}} & M_{\text{2}} & 2 M_4 & 0 & -i M_{\text{5}} & -M_{\text{5}} \\
 \f^1&0 & iM_{\text{G2}} & i M_{\text{5}} & -i M_{\text{5}} & M_3 & 0 \\
\f^2 &- iM_{\text{G2}} & 0 & -M_{\text{5}} & -M_{\text{5}} & 0 & M_3 \\
\end{array}
\right)
\la{chargemassmat}
\ee
where
\be
M_1= \fr{g_1}{2 \sqrt{2}} M_H
;\;M_2= \fr{g_2}{2 \sqrt{2}} M_H
;\; M_5= g_5  M_H
;\;M_{\text{G2}}\to \frac{g_2 M_G}{\sqrt{2}}
\ee

The product of this with its \CC\ yields the Mass Squared matrix for the charged Fermions in the Gauge/Higgs sector. It has three double eigenvalues, two corresponding to the mass of the charged vector Boson $W^+$, and one corresponding to the mass of the charged Higgs $H^+$.

\refstepcounter{orange}\la{numbers}
{\bf \theorange}.\;{\bf Fitting the Model to Experiment:} 
This \HCM\ Model has five parameters in the Action, namely the dimensionless coupling constants $g_1,g_2,g_5$
and the two masses $M_3,M_4$.  From these we have shown that there is a pair of 
 VEVs  with magnitudes $ M_G ,M_H$ satisfying the equations:
 
\be
M_G =\fr{- M_4 }{g_5}
\Lra
g_5 M_G = - M_4 \ee
\be
 M_H^2= \fr{-M_3 M_G}{ g_5}= \fr{M_3 M_4}{ g_5^2}
\Lra
 M_H^2g_5^2 =  M_3 M_4 
\la{m0g4m3m4}
\ee

So these two VEV magnitudes are determined by the parameters  $M_3,M_4,g_5$.

In fact we note that
\be
g_5^2 M_G^2 =  M_4^2 ; M_H^2g_5^2 =  M_3 M_4 
\Ra 
\fr{M_G}{M_H} = \sqrt{\fr{M_4}{ M_3}}
\ee

By looking at the pure field part of the Action after the shift of the fields that have 
VEVs we found  the following masses for the Z, the W and the neutral Higgs:

\be
 M_Z^2= \frac{1}{4} \left(g_1^2+g_2^2\right)  M_H^2
\la{MZeq2} 
\ee
\be
M_W^2=\frac{1}{4} g_2^2  
 M_H^2+ \frac{1}{2} g_2^2  
 M_G^2 
 \la{notstandard}
\ee
\be
M_{\rm H0\;Light}^2=
\frac{1}{2} \left(4
M_3 M_4
 +M_3^2-\sqrt{M_3^4+8
   M_3^3 M_4}\right)
\ee
Note that  (\ref{notstandard}) 
 is not quite the same as the \SM, because the formula there has the form
\be
{\bf Standard\; Model}: M_W^2=\frac{1}{4} g_2^2  
 M_H^2  
 \la{standard}
 \ee
without the term $+ \frac{1}{2} g_2^2  
 M_G^2 
$ that is contained in (\ref{notstandard}). This term $+ \frac{1}{2} g_2^2  
 M_G^2 
$  needs to be quite small to be consistent with experiment, and that is what gives us large masses for the two new Higgs in this Model. 
In the Standard Model, and in this \HCM, the parameter $g_2$ can be determined from the formula 
\be
\fr{G_W}{\sqrt{2}}= \fr{g_2^2}{8 M_W^2}
\ee
where $G_W$ is the Fermi constant, and $M_W$ is the W mass.
We can get the electromagnetic charge from 
the usual formula
 for the fine structure constant, except that we acknowledge the running of the coupling constant 
 and use the value appropriate for $Q^2=M_W^2$, which is :
\be
e_1 = \sqrt{\fr{4 \pi}{128 }} 
=0.313329
\la{128number}
\ee
Note that we are using this value taking into account the running of the coupling constant to $Q^2 = M_W^2$, as quoted in  \ci{PDGnumbers}.  This makes all the difference for this Model.
Note that we do not use the value for  $Q^2=0$ here.  If we did then we would get
$e_0 = \sqrt{\fr{4 \pi}{137 }} \approx .302$ , and no solution for this \HC\ SSM Model is possible with this value.  We would get an imaginary value for $M_G$, which is nonsensical.  This is the origin of the very high mases for the predicted Higgs particles, and also the reason that we only have three significant figures (if that!) for those masses.

Then the parameter $g_1$ is related to the electric charge as in the standard Model by 
\be
  \sin \q_W =\fr{e_1}{g_2}
\ee
and we get:
\be
  \sin \q_W =\fr{e_1}{g_2}
=\fr{0.313329}{0.652954}
\ee
Then we have
\be
\q_W= 0.500499
;\; \tan \q_W = \tan (0.500499)=0.546951
=\fr{g_1}{g_2}
\ee
and then 
\be
g_1
=0.546951 g_2
=(0.546951 )(0.652954)
=0.357134
\ee

Now we are ready to make the predictions for the two masses that are determined by the parameters in this Model.  They are
\be
M_{\rm H0\;Heavy}=
\sqrt{\frac{1}{2} \left(4
M_3 M_4
 +M_3^2+\sqrt{M_3^4+8
   M_3^3 M_4}\right)}
\ee
and
\be
( M_H^+)=
 M_3+2
   M_4 
   \ee
They are not quite equal, but very close and the inaccuracy of our knowledge of the number in equation
(\ref{128number}) far exceeds the difference.  They are both approximately 13,400
GeV according to this calculation.

The above results and their numerical values are summarized in the following tables:

\be
\begin{array}{|c|c|c|c|c|c|}
\hline
\multicolumn{6}{|c|}{\rm  Table (\ref{numberstable1new})\;   Experimental\;and\;Derived \;Values\; of\; Pure\; Numbers}\\
\hline
e_0= \sqrt{\fr{4 \pi}{137.03599974}}  
&e_1= \sqrt{\fr{4 \pi}{128}}  
& \theta _{\rm W} & g_{1} &
   g_2 &  g_5\\.302822 &
0.313329
& 0.500499 &
0.357134
 & 0.652954  &3.74157\\
\hline
\end{array}
\la{numberstable1new}
\ee

\be
\begin{array}{|c|c|c||c|}
\hline
\multicolumn{4}{|c|}{\rm  Table (\ref{numberstable2})\; Experimental  \;Values \; of \; Masses}\\ \hline\hline
\multicolumn{3}{|c||}{\rm Known\;
Weak\;Bosons\;(GeV) }&
\multicolumn{1}{|c|}{\rm Fermi\;Constant 
\;(GeV)^{-2} }\\
\hline
M_Z &
   M_{\text{\rm H0\;Light}} &  M_W  &  G_F    \\
  91.1876 & 125.7 &
  80.385 &
  1.1663787 \times 10^{-5} 
\\
\hline
\end{array}
\la{numberstable2}
\ee

\be
{
\begin{array}{|c|c||c|c||c|c|}
\hline
\multicolumn{6}{|c|}{\rm  Table (\ref{numberstable3})\;New\; Derived\;Values \; of \; Masses}\\ \hline\hline
\multicolumn{2}{|c||}{\rm VEVS \;(GeV) }&
\multicolumn{2}{|c||}{\rm Mass\;Terms\;in\;Action\;(GeV) }&
\multicolumn{2}{|c|}{\rm Predicted\;Higgs 
\;(GeV) }\\
\hline
M_G &
   M_H & M_{3} &
   M_{4}& M_{\rm H0\;Heavy} &
   M_{H+} \\
  16.9571
&245.049&13,249.8 &63.4463
&13,375.5&13,376.7\\
\hline
\end{array}
\la{numberstable3}
}\ee

\refstepcounter{orange} \la{lepsandquarks}
{\bf \theorange}.\;{\bf  The  Lepton and Quark Sector does not need Scalars:}
To describe the Leptons and Quarks, the best result seems to be available if one uses \UCM s. These render spontaneous breaking of SUSY unnecessary in this sector, because they remove the Squarks and Sleptons (the `Smatter') entirely from the physical fields of the theory. 

It is notorious that this removal is not much of a  loss to the theorist.
These Scalar partners of the Leptons and Quarks  make the theory much more complicated, and they are responsible for most of the 124 parameters in the MSSM, as well as the hidden sector and the messenger sector, the lack of natural suppression of flavour changing neutral currents, and the \ccp.   Also, after considerable work, the Squarks and Sleptons have not been found.  

To remove the Smatter, one starts with the usual Left SU(2) Doublets  for the Quarks and Leptons, familiar from the usual SSM, coupled  to the Higgs Multiplets $H_L, H_R$  and  the right handed Quark and Lepton SU(2) singlets through superpotential terms, with the CKM matrices, and their Leptonic versions, for flavour coefficients, in the usual way. 
 The VEVs of the $H_L, H_R$ Higgs used above in Equation (\ref{HVEV}) will give masses to the Quarks and Leptons in the usual way. The $S^a$ Higgs cannot couple to the Matter of course.

There is no present need to make this paper longer by writing this out explicitly.  The Action for the Chiral Left handed Doublets of the Quark or Lepton are the same as in \pgh\ \ref{HLD} above, with suitable renaming, and three flavours, and a coupling to colour $SU(3)$ for the Quarks.  The Chiral Right handed singlets are similar but do not couple to $SU(2)$. 

The only new thing here is that  we  implement the appropriate \CTR\ to make these into \UCM s, and simultaneously make the Higgs into \HCM s as discussed above. These \UCM s are simpler than the \HCM s and both of these were described in  \ci{four}.
 This  removes all the Squarks and Sleptons from the Field content of the 	theory. But they are still present as Zinn Sources in the theory, and the Zinn Sources $\G$ for the variation of the Squarks and Sleptons have become `Antighost Fields' $\h$.  This is still an exactly supersymmetric theory, and it is not possible to remove the complicated terms in the Zinn Action or the Antighost terms\footnote{These Antighost terms do not seem to play much of a role in the Feynman expansion. There are of course plenty of terms that arise in the Feynman expansion from the Zinn Sources, since they are coupled to the Gauge Fields and the Higgs Fields. All of these  are needed to satisfy the \PB, of course.}. 

Clearly this makes the Quantum Field part of the Lepton and Quark Sector of this kind of Model  nearly identical to what it is in the \SM\footnote{The $SU(3)$ Gluons and Gluinos also are a puzzle.  Since there are no Squarks, the gluinos have nothing to attach to except other Gluons and Gluinos.  Does SUSY need to be split in this sector?}.

\refstepcounter{orange} \la{gaugehiggsumm}
{\bf \theorange}.\;{\bf The  Gauge/Higgs Sector does need some Scalars:}
The situation is somewhat different for the Higgs Chiral Scalar Multiplets.  As we saw above, here we do need to have Scalars, because they are needed to give rise to \GSBS.  There is an advantage in using the \CTR\ to remove half of the Scalars from the Higgs Chiral multiplets to make \HCM s. These smaller \HCM s  ensure that the Z Boson is not degenerate with a neutral Higgs Boson, and that the W Boson is not degenerate with a charged Higgs Boson. Their VEVs  also give masses to the Leptons and Quarks.

But no Scalar partners are needed for the Quarks and Leptons.  So it is easy to account for the successes of the non-supersymmetric \SM, because the Squarks and Sleptons are simply absent from this new SSM.

When we use these `\HCM s', the particles turn out to still have some Fermi-Bose degeneracy after gauge symmetry breaking, as was seen in the long analysis above.

\refstepcounter{orange} \la{conclusion}
{\bf \theorange}.\;{\bf Conclusion:}
The Model presented here has the particles of the usual \SM, plus two very heavy Higgs with mass 13.4 TeV, plus superpartners for the particles in the Gauge Sector and the Higgs sector. 
These are `Gauginos and Higgsinos'.  The  Gauginos are the Photino ${\widetilde \g}$, the Zino  ${\widetilde Z}$  and the Wino  ${\widetilde W}^+ $.  The Higgsinos are  the  ${\widetilde H}_{0 \rm  \; Light}$,  ${\widetilde H}_{\rm 0 \; Heavy}$ and ${\widetilde H}^{\rm +}$.   Each of these has the same mass as its superpartner and we wrote down the Fermionic mass matrices in equations (\ref{neutmassmat}) and (\ref{chargemassmat}) above.  Here is a summary:

\be
\begin{array}{|c|c|c|c|c|c|c|}
\hline
\multicolumn{7}{|c|}{\rm This\; model \; has \;the\; Standard \;Model \;For \; Fermionic\; Quarks \;and \;Leptons\; \;plus}\\
 \hline
\multicolumn{7}{|c|}{\rm Degenerate\; Masses \;for \;the\; Gauge \;and\; Higgs\; Sector \;With\;Masses\; (GeV)}\\
\hline
{\rm Bosons} &{\rm Photon} \; \g &{\rm W^+}  &{\rm Z^0}  &{\rm Higgs\; H^0}   &{\rm H^0_{\rm Heavy}}  &{\rm H^+_{\rm Heavy}} \\
\hline
{\rm Deg. \;Free.}\;&2&6&3&1&1&2\\
\hline
{\rm Fermions} &{\rm Photino} \; {\widetilde \g} &{\rm Wino \;{\widetilde W}^+}  &{\rm Zino\;{\widetilde Z}^0}  &{\rm Higgsino\; {\widetilde H}^0}   &{\rm  {\widetilde H}^0_{\rm Heavy}}  &{\rm  {\widetilde H}^+_{\rm Heavy}}\\ 
\hline
{\rm Deg. \;Free.}\;&2&8&4&2&2&4\\
\hline
{\rm Mass}&0& 80.385 & 91.1876 & 125.7 &\approx 13,400&\approx  13,400\\
\hline
\end{array}
\la{sumtable}
\ee
The number of degrees of freedom (denoted $\rm Deg. \; Free.$ in Table (\ref{sumtable})) can be deduced from the detailed work on the masses earlier in this paper.  For example, note that the Zino has two Weyl Fermions, from Equation (\ref{neutmasseig}), and each of them has two Degrees of Freedom.  These numbers arise because the Gauge Multiplets do not have suppressed SUSY charges, but all the Higgs Multiplets are \HC.

Aside from these \GaH, and the extensive Zinn sector and the Antighost sector\footnote{There are also antighosts and ghosts of the usual kind which are needed as Faddeev Popov Ghosts for the various gauge theories in this model when they have been gauge fixed.  These do not seem to couple to the Antighosts that appear  in the \CTR s, even when the whole Action is coupled to \SG.  These \CTR\ Antighosts seem to be quite dormant, but that certainly needs more investigation.} which do not cause any physical effects, the Model is nearly identical to the usual \SM\footnote{It is not quite identical because of the difference between  equation (\ref{notstandard}) and equation (\ref{standard}).  This difference is small and it is also the reason for the large new Higgs masses.}, except that it has two extra very heavy Higgs Bosons. That is encouraging, because it is consistent with the precision tests of the \SM\ that are continually increasing\footnote{See for example \ci{PDGstandardModel}}.   And to get to this point no spontaneous or explicit breaking or extra hidden sector of SUSY was needed, and SUSY is still exactly present.

We cannot avoid these \GaH, because they arise naturally from the need to implement \GSBS, with the observed masses for the observed Gauge Bosons and Higgs.  	What should be done with these?
Can we keep them as part of the predictions of this theory?
Perhaps that is not inconceivable, because they are really quite well hidden experimentally.

In the first place, note that the absence of the Squarks and Sleptons means that these \GaH\ do not couple to the 
Quarks and Leptons at all.  So they cannot be produced by the particles in the Matter Sector, and they cannot decay into those particles either.
All the \GaH\ would need to be produced in pairs, because this Model does respect R-parity.    The Wino ${\widetilde W}^- $ is the most obvious such particle to look for, since it is charged.  The most obvious production mechanism would be pair production of two winos ${\widetilde W}^++ {\widetilde W}^- $ in $e^+ e^-$ collisions, and this would be mixed in with W Boson pair production at exactly the same energy.  Another production mechanism is 
\be
W^+ \ra {\widetilde W}^+ + {\widetilde \g}
\la{wdecay},\ee
but this cannot happen on shell. And then the Wino has to decay back into the W, presumably, in order to decay into Matter. 
All of this requires careful analysis.

Assuming that there are no special relations or consequences among the couplings here,
which is not obvious, it is to be expected  that the only 
 stable Gaugino or Higgsino will be the  Photino, which is massless.  There are lots of infrared issues here, similar to those for the Photon in the \SM\ or Quantum Electrodynamics.  Those gave rise to the Block-Nordsieck 
treatment \cite{Stirling:2012ak}, and these Photinos look like they might have similar issues.

This Photino is probably not a satisfactory candidate for dark matter, because it is hot dark matter, always relativistic, and studies indicate that cold dark matter is needed to account for the observed dark matter \ci{darkmatter}. That may be a serious problem  in this Model.  On the other hand, it is conceivable that a detailed analysis of this model, in terms of mass and charge eigenstates, might come to a different conclusion.

 An off-shell Wino could decay to a photino (invisible) and a W Boson in a   process like (\ref{wdecay}).  An infrared photino in that process really takes very little with it besides some spin. The rest of the \GaH\ are quite hidden, because they are neutral, except for the very heavy Higgsino ${\widetilde H}^{\rm +}$. And of course all of them are completely decoupled from the Quarks and Leptons.  Their only possible decay mechanism is through the Gauge and Higgs Bosons, which are degenerate in mass with them, ultimately leaving Matter, Photons and invisible Photinos, and perhaps it is easy to confuse them with other processes. 

So at the most superficial level, it seems that this model does not actually have any obvious particles beyond those of the \SM, except the very massive new Higgs Bosons that are probably out of the range of current experiments. But the author does not pretend to have analyzed the consequences of  these \GaH, or to be able to tell whether these \GaH\ have already been excluded by experiment, particularly since the experimental analyses would have been  based on the SSM as usually formulated, which is completely  different from the present Model.  
More work is needed here. Perhaps this Model can be easily proved to be inconsistent with experiment, since it is very constrained.

  It is possible to shift some of the masses of these superpartners by adding more \UCM s to the Gauge/Higgs sector of this Model, and there are lots of issues to be considered, and no more will be attempted in this paper.  Note however that adding \UCM s will not change the Boson masses, and it will not vary the prediction of the two new Higgs masses.  This is a consequence of the fact that \UCM s have only Fermions, and no Scalars.

 Many new experimental results are coming out, and they tend to confirm that the \SM\ is still looking healthy \cite{Aad:2016zqi,Aad:2016naf,Aaboud:2016dig,Aad:2015tna,Khachatryan:2016mud}.  This of course is consistent with the absence of Squarks and Sleptons and the close fit of the present SUSY SSM to the old \SM.

Perhaps it is  not necessary to hypothesize Sleptons or Squarks, because they have not been found, and the simpler Model in this paper is available, and it does not have the serious problems mentioned earlier.  But SUSY is complicated, and this model may have other serious problems that have not yet been noticed by the author.

The Model raises plenty of other questions of course.
Is it possible, or necessary, to modify this Model, by adding \UCM s, so that it is consistent with the known experimental evidence? Is it inconsistent with the known results in some way?
  Is there a viable explanation for dark matter here?  Do the non-renormalization theorems still apply in the absence of the supercharges that arises here?  Does the vacuum energy remain zero at loop orders, so that the cosmological constant problem \ci{weinbergcosmo} really does not arise at all?  Are there anomaly problems or parity assignment problems here?  	Can they be resolved?  Does this theory have some other kind of problem? 
 What about CP violation? Does the Gravitino remain massless?  Is there any problem coupling this Model to Supergravity?

\begin{center}
 {\bf Acknowledgments}
\end{center}
\vspace{.2in}

  I thank  Carlo Becchi, Friedemann Brandt, Cliff Burgess, Philip Candelas, Rhys  Davies, James Dodd, Mike Duff,  Chris Hull, Pierre Ramond, Peter Scharbach,   Kelly Stelle, 
J.C. Taylor and Peter West for stimulating correspondence and conversations.  Raymond Stora introduced me to spectral sequences, which eventually brought  \cdss s forward, and he will be missed very much. I particularly want to thank 
Xerxes Tata who took the trouble to insist on the importance of the tachyon problem for the \cdss, and on the issue of the non-conservation of the Supercharge. These were the origin of the \HCM, which gave rise in turn to the \CTR, which is the foundation of this paper and its companion on the SSM. Some of this work was done at the Mathematical Institute at Oxford University, and  I thank Philip Candelas and the Mathematical Institute for hospitality.

\tiny \articlenumber
\\
\today

\begin{thebibliography}{99}
\bibitem{majunchiral}  
Equation (74) of \ci{four} sets down the \PB\ for the Majorana \UCM.   The \PB\ for the Dirac \UCM\
works in a similar way, but for the Dirac case, which arises out of Section 37 of \ci{four}.
Equation  (82) is half way there.
\end{thebibliography}
\end{document}